# The rejection that defies anti-rejection drugs.
## Chronic vascular rejection (allograft vasculopathy): The role of terminology and linguistic relativity

Vladimir M. Subbotin and Michael V. Subotin

Madison, WI, USA,

Solid organ transplantation has by now become a common medical procedure. Owing to the introduction of new immunosuppressive drugs, the allograft loss due to acute rejection has been reduced significantly over time. Tragically, the number of donor organs lost due to allograft vasculopathy (AV), generally named "chronic vascular rejection" or "chronic rejection", has remained significant and unchanged for decades. We argue that designation of AV as chronic rejection, and its classification as a delayed long-lasting reaction of recipient immune effectors against donor alloantigens have given us the wrong impression that we have identified the necessary cause of the disease. However, whatever treatment options we have in the anti-rejection toolbox, despite their success in treating acute rejection, do not work for AV. Yet, the scientific community has continued to conceptualize and approach AV within the alloimmunity and rejection model. Due to unproductive research from the alloimmunity perspective, the number of transplanted hearts lost due to this pathology today is almost the same as it was fifty years ago. We believe that this phenomenon falls under the rubric of linguistic relativity, and that the language we chose to name the disease has restricted our cognitive ability to solve the problem. While the initial perception of the AV as chronic rejection was logical and scientific, the subsequent experience revealed that such perception and approach have been fruitless, and likely are incorrect. Considering our tragic failure to prevent and treat allograft vasculopathy using all available knowledge on alloimmunity and rejection, we must finally disassociate the former from the latter. A good way to start this process is to change the words we are using; particularly, the words we chose to name the disease. We have to step out of the alloimmunity rejection box.

Let's pause between our words,
Speak and fall silent again,
So that the meaning of the word just spoken,
Sounds a clearer echo in our heads.
Let's pause between our words.

Andrey Makarevich[1]*

## Introduction[1]

Solid organ transplantation across the allogeneic barrier, pioneered by Thomas Starzl, has by now become a common medical procedure. Surgical challenges, although they were considerable, were soon overcome for solid organ transplantation, e.g., for kidney transplantation[2], and later also for transplantation of other organs.

The main obstacle of this innovative treatment was foreseen early and analyzed in particular detail in the celebrated studies by Sir Peter Medawar and Sir Frank Burnet[2, 3]. Long before organ transplantation became common procedure Peter Medawar writes:

"There is only one reason, an immunological reason, why a homograft

---

should fail where an autograft would have succeeded: the rejection of homografts is a reaction of the same general kind as that which is responsible for the protection of the body against bacteria, viruses and other parasitic organisms" [2].

Homograft rejection was well understood much earlier than organ transplantation became common, and the nature of this phenomenon was investigated from both biological and medical perspectives. The first clinical analysis of homograft rejection was published in 1956[4] and the term "rejection", which literally reflects rejection of the donor tissue by the host, became universally accepted with clear reference to immunologic nature of the process.

Initially the term "rejection" was used without specifying the time of onset. The terminology "acute rejection"[5] and "hyper acute rejection"[6], the latter referring to preexistent antibodies, appeared in publications about ten years later.

The major milestones in the history of development of anti-rejection drugs, or immunosuppression therapies, are a testimony to the fantastic successes of numerous scientific studies. The history of development of anti-rejection drugs also illustrates how science works, or should work: the scientific community encountered a medical problem (rejection), investigated its nature (the host's immune reaction to donor alloantigens), and developed effective procedures and drugs that suppress immune reactions against donor alloantigens.

This triumphal textbook story of how science prevailed in the difficult task of vanquishing rejection has one counter-narrative: an unpreventable and mostly untreatable complication which kills transplanted organs at a much later time, ranging from months to several years after transplantation (up to 20 years). In the first two reports in 1965 this pathological condition was named "chronic vascular rejection"[7, 8] and classified as a delayed long-lasting reaction of recipient immune effectors against donor alloantigens. Today, medical literature often omits word "vascular" and simply uses the term "chronic rejection", but these two terms are interchangeable and equally common, e.g.,[9, 10]. However, let us consider this situation from a linguistic perspective.

## Organ rejection after allotransplantation: hyperacute, acute, and chronic

It was already known for a hundred years that a preexisting antigen could cause immediate severe reactions in blood transfusion, since the ABO blood group antigens are the most immunogenic of all the blood group antigens. Hence, in organ transplantation, the severe reaction that is caused by the presence of antidonor antibodies in the recipients' blood before transplantation and ultimately leads to graft loss, was termed hyperacute rejection, with an obvious reference to the immunologic nature of the reaction[11,12]. Today, hyperacute rejection due to ABO incompatibility is an avoidable and extremely rare immunologic complication. However, even if such donor-recipient ABO conflict is unavoidable, the severe complications are preventable by certain procedures within the immunologic toolbox, e.g.,[13]. There are currently many other options available to predict the likelihood of acute donor-specific antibody-mediated responses, which include flow crossmatching and the 'virtual' crossmatch; these options were developed based on the concepts of alloimmunity and rejection[14].

Then, with a different time of the onset, acute rejection comes into play. Even with a matching ABO blood group and best HLA matching, solid organ transplantation across the allogeneic barrier inevitably results in donor organ rejection by the recipient. Acute rejection occurs days after organ transplantation, though it can sometimes occur earlier or later[15]. The immunologic nature of this pathologic condition became absolutely clear when Sir Peter Medawar and Sir Frank Burnet published their research[2,3]. Again, since antibody- and cell-mediated immunological damage was studied in detail (for a review see[16,17]), numerous preventive and treatment options were investigated and clinically applied, e.g., antiproliferative agents, steroids, polyclonal anti-T cell antibody, Cyclosporine A, anti-CD3 monoclonal antibody, FK-506 (Tacrolimus), anti-IL-2R monoclonal antibody, and Sirolimus (for comprehensive reviews see[18-21]).

Again, this triumph was due to brilliant scientific and clinical investigations within the immunologic model of the pathology.

Finally, the term "acute rejection" provided foundation for new medical terminology when deterioration of donor organs was found to occur much later, years after transplantation, e.g.,[22,23]. Naturally, considering the presence of donor alloantigens and persistent activation of host immune responses against donor tissues, the scientific community likewise perceived this problem as a persistent, long-term allograft rejection and hence studied it within the immunologic model. When it came to naming this newly discovered complication, the choice appeared obvious. It was named "chronic vascular rejection" or simply "chronic rejection", terms which at the time appeared intuitively compelling and scientifically sound.

It could be argued that judging scientific understanding of the past based on today's knowledge is not appropriate. Nevertheless, it seems odd to us that nobody at that time (40-50 years ago) was surprised by the peculiar microscopic pathology of this newly discovered delayed rejection complication: while donor cell damage by recipient immune effectors, which was familiar from observations on acute rejection appeared to be very subtle or sometimes not present at all, the new pathologic condition consisted of uniform cell proliferation in arteries of transplanted organs, primarily affecting one compartment of arterial wall – the *tunica intima*; this progressive cell proliferation creates an excess of neointimal tissue which obstructs the arterial lumen, causing graft ischemia and loss. The following concise and widely accepted definition of chronic rejection comes from the research of Pittsburgh transplantation scientists, the most advanced group in this field:

"Chronic rejection (CR) primarily manifests itself as a progressive obliterative arteriopathy (OA) that most frequently affects medium-sized muscular arteries of vascularized organ allografts. The lack of similar changes in isografts argues strongly for an immunologic basis."[24]

"CR can be broadly defined as a largely indolent, but progressive form of allograft injury characterized primarily by fibrointimal hyperplasia of arteries, or obliterative arteriopathy (OA), interstitial fibrosis and atrophy of parenchymal elements."[25]

The other technical names of chronic rejection or transplant obliterative arteriopathy include: allograft vasculopathy, transplant arteriosclerosis, allograft arteriosclerosis, transplant vascular sclerosis, graft coronary artery disease, post-transplant vasculopathy, transplantation vasculopathy, transplant neointimal formation, transplant intimal hyperplasia, allograft intimal hyperplasia, allograft intimal proliferation, allograft intimal thickening, transplant intimal thickening, transplant fibromuscular dysplasia, allograft fibromuscular dysplasia, allograft arteriopathy, transplant arteriopathy, etc. However, there is one widely adopted definition for this pathology, which serves as the name of the disease, as its classification, and as a theory regarding its nature: it is the original term "chronic rejection". All the technical names listed above refer, directly or indirectly, to alloimmune nature of the pathology. Even though one author introduced the somewhat neutral term "graft vascular disease" (GVD)[26], trying to avoid unsubstantiated assumptions about the underlying pathogenesis, the notion of continuous recipient immune attack against donor alloantigens, i.e., rejection in its literal sense, still holds sway in the scientific community. It should be mentioned that some analyses of chronic rejection refer to its mechanisms as "multifactorial", which alludes to nonimmunological causes e.g.,[27]. However, priority is given to the "alloimmunity and rejection" model (for review see[28,29]). Alternative conceptions of the pathogenesis of allograft vasculopathy are very rare[30, 31] and receive very little attention.

From the standpoint of its pathology, allograft vasculopathy always forms patterns and this is considered to be a diagnostic landmark of chronic rejection[32,33]. Transplant arteriosclerosis affects donor heart, pancreas, kidney, bowel and vascularized composite allotransplantations, and also, though much more rarely, transplanted liver[34, 35] and lungs[33, 36-38].

Therefore, summarizing the above, we can conclude: chronic rejection consists of two types of tissue pathology: 1) some features of acute rejection, which are not necessarily present, and 2) features peculiar to chronic vascular rejection, which are always present. A comparison of these two types of tissue damage in chronic rejection suggests an inevitable conclusion: the first type of damage, which is due to chronic immune attacks against alloantigens (by cellular, antibody, or complement components), followed by necrosis and fibrosis, e.g.,[39,40], is predictable and scientifically logical. On the contrary, the second type of chronic rejection pathology involves no or minimal tissue damage that could be blamed on chronic immune attacks, but always manifests itself by persistent, uniform proliferative morphogenesis in the arterial *tunica intima* of transplanted organs[32,33], which is counterintuitive and perplexing.

## Chronic rejection: stalled progress in prevention and treatment and a puzzling trend in research publications

The tremendous therapeutic success in curbing acute rejection by a newly discovered class of drugs, collectively known as immunophilins, was originally anticipated to be extended to the chronic rejection[41-45], however, this prediction, as we know by now, fell short of expectations.

Most importantly, whatever treatment options we have in the "anti-rejection toolbox", despite their success in treating "classical" acute rejection, do not work for chronic rejection. The bibliography for this statement it too extensive to cite here: a vast array of factors that could be considered as part of immune regulations/effectors, or even remotely associated with them, has been proposed as a possible cause of transplant arteriosclerosis and thoroughly tested, but all these attempts have failed to produce clinically viable results. These attempts at prevention and treatment cover everything from large domains such as innate and adaptive immunity, cellular and antibody-mediated immune responses, to smaller domains such as soluble and membrane-associated antigens, complements, etc. Everything that has been suggested as a cause of chronic vascular rejection within immunological models was countered with the appropriate treatment, but all have failed clinical tests. Therefore, article titles like "Chronic rejection—an undefined conundrum"[46], "Allograft vasculopathy: the Achilles' heel of heart transplantation"[47] or "Chronic Rejection: An Enduring Enigma of Transplantation Biology"[48], and a notable contribution by Dr. Starzl "Organ Transplantation: A Practical Triumph and Epistemologic Collapse"[49] are familiar to the readers. Yet, the scientific community has continued to conceptualize allograft vasculopathy within the alloimmunity model over the past decades[9,50], and up to the present day[51-53].

Surprisingly, a failure to solve the allograft vasculopathy complication within the alloimmunity model has coincided with a disconcerting trend in research publications related to chronic vascular rejection and allograft intimal hyperplasia. Considering the magnitude of the problem, it would be logical to expect an increase of basic research on allograft vasculopathy pathogenesis, and mechanisms driving post-transplant neointimal formation, materializing in a corresponding increase of the relevant scientific publications. However, PubMed searches revealed that while the number of publications on organ/tissue transplantation procedures has rapidly increased, the number of publications on mechanisms of chronic vascular rejection, and neointimal formation has declined. This decline is evident by numerous PubMed searches using all possible alternative terms for "chronic vascular rejection" "allograft vasculopathy" and "allograft intimal hyperplasia" (Figure 1). Recently, repeated PubMed searches confirmed the above results (data not shown).

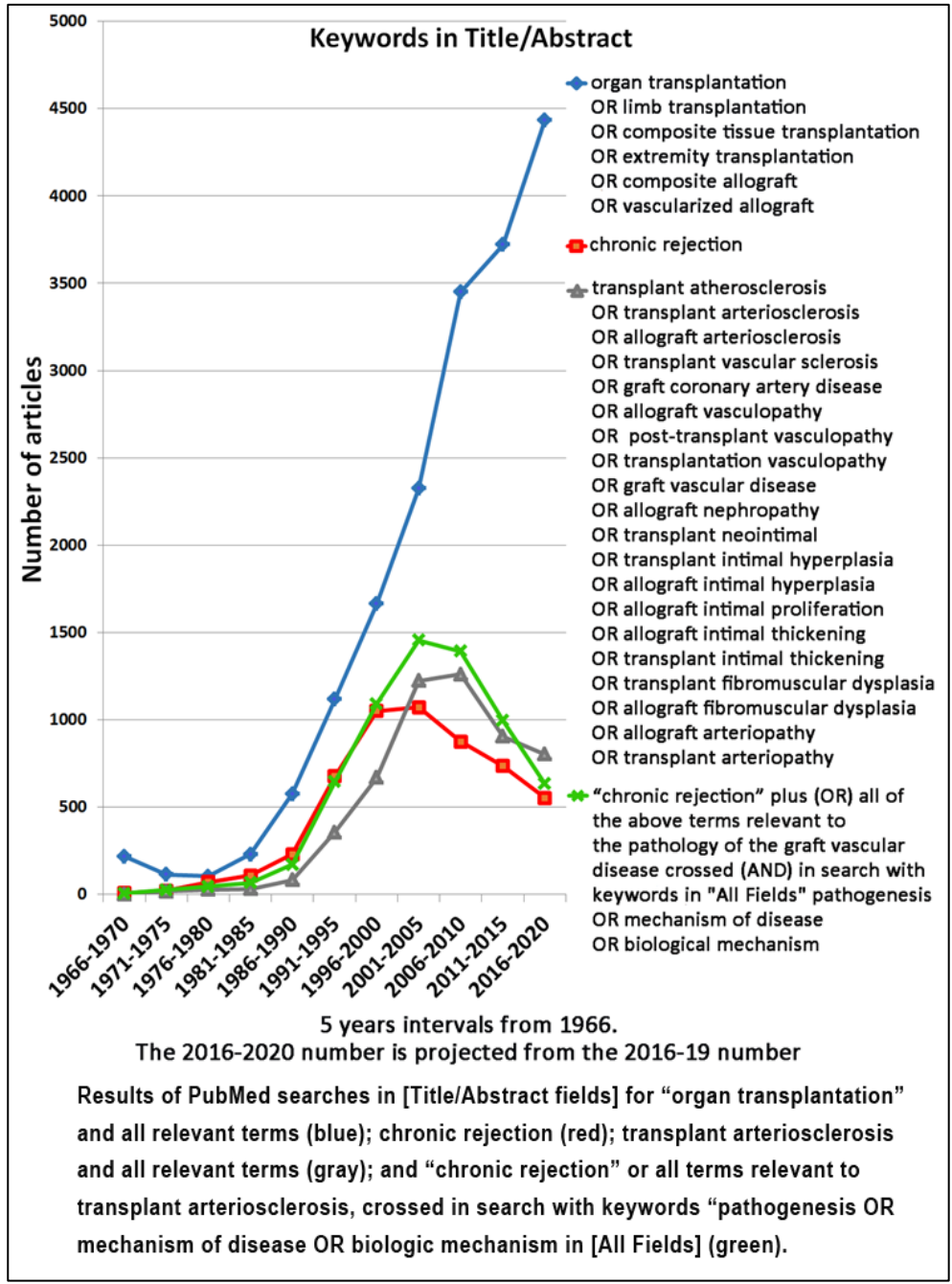

Figure 1. Reproduced from[54] with permission of the publisher.

Therefore, assuming that the specific keywords in bibliographic records of scientific publications reflect the study aims and design, the PubMed searches have shown that we study and publish more and more on organ/tissue transplantation procedures, which are all affected by chronic vascular rejection, while studies of chronic vascular rejection and allograft intimal hyperplasia itself have declined[54]. Thus, we are facing a growing volume of this complication whose pathogenesis is not understood, making allograft vasculopathy untreatable, while at the same time failing to make progress in its research.

Can we conclude from this trend that we have given up and the disease can declare victory? We do not want to entertain an affirmative answer to this question, because the outcome would be tragic and unacceptable. Instead, as the poet suggests, let us "pause between our words"[1] and seek a possible explanation for this dire situation.

## To put it simply: a conversation with a car mechanic

First, let us simplify this complex situation. We invite our readers to zoom out from the numerous intricate immunologic details and look at the situation as a lay person would. For example, let us imagine that we want to explain the problem to a car mechanic, while he or she is working on our car. Our narrative would be:

"In some people a vital organ can fail. The rest of the body of these patients is just fine, except a particular organ, such as their heart. The only way to save these patients is to give them new organs from donors; we call this procedure organ transplantation, and it has become a common medical procedure. Even before organ transplantation became available, we knew from numerous experiments that the immune system of a patient who receives an organ from a donor would fight against it by killing donor's cells, unless the recipient and donor are close relatives, ideally twins. However, due to practical reasons, most patients can only receive donor organs from non-related persons. Then, their immune system recognizes donor organs as foreign cells and starts to kill them, the same way as our immune system fights against microbes. We named this immune reaction against donor organs "acute rejection", and it occurs days or weeks after a transplant. Fortunately, scientific progress gave us certain measures against this type of organ rejection, primarily in the form of drugs that effectively suppress the immune system and protect donor organs from damage. Thanks to these treatments, acute rejection became controllable, and the transplanted organs became tolerated or even accepted like the patient's own organs.

However, in a significant number of patients transplanted organs begin to deteriorate much later, years after transplantation, and we called this delayed pathology "chronic rejection". This chronic rejection comprises two types of pathologic manifestations. The first type involves damage of donor's cell caused by the recipient's immune system, which is what we would expect. The second type consists of a surprising cell growth inside arteries of transplanted organs, which constricts the arteries and blocks blood supply to the organ. The first type of chronic rejection could be treated by the same drugs that effectively control acute rejection. However, the second type of chronic rejection, the cell growth inside arteries, does not react to these drugs at all. It is really frustrating, because, in spite of all our efforts, the number of transplanted hearts lost due to this chronic rejection today is almost the same as it was fifty years ago.

For good measure, we could add: "since we haven't found an effective way to treat this chronic vascular rejection, we decided not to dwell too much on this deadly pathology in our research and publications".

We can imagine that at this moment the mechanic would stop working on our car and ask the following questions: "Guys, then why have you decided that this mysterious delayed cell growth that clogs up arteries is a form of organ rejection by the immune system? You just said it does not look like the kind of rejection you can treat and does not react to the drugs you use to treat rejection, so why do you think it works in a similar way? If it doesn't walk like a duck or quack like a duck, why should it be a duck?"

What could we say in response? Perhaps something like the following: "Because we expect that the immune system of a recipient rejects donor organs, we associated this delayed arterial cell growth inside arteries with immune reaction and perceive it as a manifestation of chronic rejection. Now we are recognizing that the name "chronic rejection" is a poor fit for this condition, but we have become used to this term and continue to use it."

At this juncture, we can let the mechanic continue working on our car and ask ourselves the following question: does this explanation sound logical and convincing to us? Our answer should not come as a surprise: it does not.

To further reflect on our mechanic's comments, we should ask ourselves another question: why have we named the delayed arterial pathology of transplanted organs "chronic vascular rejection"? A follow-up question inevitably emerges: does it matter how we name a disease?

**Does the name of a disease matter?**

Evidently, it does. The naming of diseases and it consequences are subject of intensive debates in medicine.

In 2015 the World Health Organization (WHO) published guidelines for naming new human infectious diseases[55]. This set of standard best practices for naming new diseases was created in close collaboration with the World Organisation (sic) for Animal Health and the Food and Agriculture Organization of the United Nations, and in consultation with the International Classification of Diseases. The main aim of the guidelines was to avoid inappropriate names, thereby minimizing unnecessary detrimental effects on nations, ethnic groups, and people in general, as well negatively affecting economies. Importance of these guidelines was further emphasized in a Science Journal publication[56].

According to these guidelines, correct use of terms for naming diseases and conditions may include generic and specific descriptive terms, the causative pathogen, if it is known, severity, seasonality, etc.[55]. At the same time, the WHO guidelines recommended avoiding inclusion of certain inappropriate terms, e.g., geographic locations, names of individuals, cultural, population, industrial, occupational, or national/ethnic references, names of animal or food, as well as terms that incite undue fear (e.g., unknown, fatal, epidemic)[55]. One paragraph of these guidelines reads: "If an inappropriate name is released or used or if a disease remains unnamed, WHO, the agency responsible for global public health events, may issue an interim name for the diseases and recommend its use, so that inappropriate names do not become established"[55]. The same authors also write in the WHO Notes for Media: "Once disease names are established in common usage through the Internet and social media, they are difficult to change, even if an inappropriate name is being used. …The best practices apply to new infections, syndromes, and diseases that have never been recognized or reported before in humans, that have potential public health impact, and for which there is no disease name in common usage. They do not apply to disease names that are already established."[57] The latter notion implies that if an inappropriate term has been used to name a disease or syndrome, or, for that matter, to name any medical entity, it may remain in use indefinitely, even if we know that it is completely inappropriate.

Consideration of these WHO guidelines suggests at least two questions worth discussing. The first one is technical: why do we use inappropriate terms for naming diseases in the first place? A plausible answer is that we use these terms because of their associatory proximity to an already known disease or pathology. In plain English, it means that we do not invest enough thought in naming a phenomenon that we have discovered and are studying; we simply pick a word that appears to be associated with it in some way, even if it is not appropriate, and the name becomes established.

The WHO guidelines rightly emphasized that giving a wrong name to a pathological condition (i.e., disease) can stigmatize particular religious or ethnic communities and create various kinds of social problems.

But are these the only obstacles that can arise from inappropriate naming? There is a view that misnaming in medicine may be associated with harm of a different nature, with even greater potential consequences: misnaming a disease could impede understanding of its pathogenesis and, consequently, make it impossible to prevent and cure it. For example, misleading terminology has been shown to hinder understanding of pathologic classification in such an important area as cancer, affecting communication between clinical groups[58].

In another case study, described in the essay "On causality and the problem of aneurysms," Martin D. Tilson (recognized world authority on aneurysms) writes:

"I believe that the term "atherosclerotic aneurysm" has been very unfortunate, because it has given us the notion that we have identified the necessary cause, the same as we have in the case of "syphilitic" aneurysm."[59].

One might object: what difference does the name make in the study of a disease? A name could be subjective or incorrect, but it is just a name, which should not interfere with scientific research, since we may suppose that the latter is by definition objective and correct. As a matter of fact, the issue is more complicated than it may seem at a glance. It has long preoccupied philosophers and scientists in various guises. The German philosopher Friedrich Nietzsche put it in striking terms late in the 19th century:

"*Only as creators*! – This has given me the greatest trouble and still does: to realize that what things *are called* is incomparably more important than what they are. The reputation, name, and appearance, the usual measure and weight of a thing, what it counts for – originally almost always wrong and arbitrary, thrown over things like a dress and altogether foreign to their nature and even to their skin – all this grows from generation unto generation, merely because people believe in it, until it gradually grows to be part of the thing and turns into its very body. What at first was appearance becomes in the end, almost invariably, the essence and is effective as such." …– But let us not forget this either: it is enough to create new names and estimations and probabilities in order to create in the long run new "things"."[60]

In the 20th century a similar notion was addressed in the hypothesis of linguistic relativity, also known as the Sapir–Whorf hypothesis.[2] This hypothesis suggests that language determines thought and linguistic categories influence and restrict cognitive

---

[2] Although it is customary to mention that Edward Sapir also wrote works which are not in favor of linguistic determinism, and that Sapir and Benjamin Lee Whorf never co-authored any works and never stated their ideas in terms of a hypothesis, Sapir is the undisputable successor of Boas and Humboldt in 20th-century American linguistics (Koerner, E.K. (1992) The Sapir-Whorf hypothesis: A preliminary history and a bibliographical essay. *Journal of Linguistic Anthropology* 2 (2), 173-198). Apart from adapting the ideas of Boas and Humboldt, in the book "Language: An introduction to the study of speech" (Harcourt, Brace and Company, page 232) Sapir writes: "Language and our thought-grooves are inextricably interwoven, are, in a sense, one and the same." The notion of "linguistic relativity", at least in its weak form, is clearly stated in Sapir's 1929 contribution "The Status of Linguistics as a Science", Sapir, E. (1929), *Language*, 207-214.

categories.[61-64]. Benjamin Lee Whorf, an American linguist who proposed what came to be known as the Sapir–Whorf hypothesis, observed and analyzed multiple examples of human behavior where names of things affected human perception more strongly than the physical nature of these things. In his work 'Relation of habitual thought and behavior to language', in the chapter 'The Name of the Situation as Affecting Behavior' Whorf writes:

"… Thus around a storage of what are called 'gasoline drums' behavior will tend to a certain type, that is, great care will be exercised; while around a storage of what are called 'empty gasoline drums' it will tend to be different - careless, with little repression of smoking or of tossing cigarette stubs about. Yet the 'empty' drums are perhaps the more dangerous, since they contain explosive vapor. Physically the situation is hazardous, but the linguistic analysis according to regular analogy must employ the word 'empty', which inevitably suggests lack of hazard. The word 'empty' is used in two linguistic patterns: (1) as a virtual synonym for 'null and void, negative, inert,' (2) applied in analysis of physical situations without regard to, e.g., vapor, liquid vestiges, or stray rubbish, in the container. The situation is named in one pattern (2) and the name is then 'acted out' or 'lived up to' in another (1); this being a general formula for the linguistic conditioning of behavior into hazardous forms."

… In a wood distillation plant the metal stills were insulated with a composition prepared from limestone and called at the plant 'spun limestone'. No attempt was made to protect this covering from excessive heat or the contact of flame. After a period of use the fire below one of the stills spread to the 'limestone', which to everyone's great surprise burned vigorously. Exposure to acetic acid fumes from the stills had converted part of the limestone (calcium carbonate) to calcium acetate. This when heated in a fire decomposes, forming inflammable acetone. Behavior that tolerated fire close to the covering was induced by use of the name 'limestone' which because it ends in '-stone' implies noncombustibility."[64].

As we can see, the name of the object or situation indeed dramatically affects human cognition and behavior. Modern scholars also share this view. Richard Wright writes on this ability of language to structure and frame phenomena we encounter "… language allows for the imposition of the structure and meaning on the world. In this sense, a sentence or set of sentences does not reflect or represent some external reality, but frames or constructs that reality". And further: "… conceptual commitments place predetermined constraints on the thinker as well as on the thoughts he or she is allowed to entertain"[65]. Linguists and other scientists in various fields continue to analyze this phenomenon, and some have expressed concerns that its practical implications have not received sufficient attention. Gary Lupyan writes:

"Much of human communication involves language. … Attempts to understand how an essentially unlimited array of meanings can be communicated using finite ordered sequences of sounds has spawned disciplines from information theory to psycholinguistics, to pragmatics. Yet, a central question concerning this fundamental property of natural language has received relatively little attention: What are the cognitive consequences of naming?"[66].

Restating our question: is it valid to compare the role that names and language play in the above situations described by Benjamin Whorf[64] to the role that names could play in perception and study of diseases? In other words, could the names we give to diseases affect our perception of diseases and our ability to find a cure, i.e., fundamentally affect the study of disease? They can indeed. Psychological and psychiatric studies have demonstrated that naming and classification of mental diseases affect the way we approach and study them[67-69].

Martin D. Tilson and coauthors found that that the same notion is applicable to non-mental diseases, i.e., to diseases that are subjects of internal medicine, for example the abdominal aortic aneurysm (AAA), which in medical practice was named as "atherosclerotic aneurysm"[70-72].

Indeed, decades of study of the lethal pathology of AAA under the name of "atherosclerotic" aneurysm did not yield any progress in prediction of its occurrence or in prevention of aortic rupture[59, 73]. However, the situation changed drastically when investigations of M.D. Tilson and coauthors ceased to be constrained by the naming and perception of the disease as "atherosclerotic". They found that there is a strong anti-sclerotic component in the pathogenesis of AAA, i.e., activation of matrix metalloproteinases in aortic tissues that degrade collagen and elastic fibers and compromise the integrity of the aortic wall[74,75]. Furthermore, the study of AAA without these pre-conceived notions dictated by its former name revealed strong immunologic and genetic components in the pathogenesis, which made it possible to predict occurrence of the disease by study of family tree genetics and to avoid fatal outcomes by detection of the silent pathology and preventive surgery[76,77].

Thinking outside the "atherosclerotic" box was the most fruitful step in the study of AAA, which was brilliantly performed by Tilson and coauthors. The departure from the widely held perception of aneurysm as an atherosclerotic disease[70-72] was intuitively compelling, logical, and scientific: atherosclerosis and aneurysm pathologies have few common features but many differing ones, a circumstance already noted by Busuttil *at al*., and Tilson and Stansel in 1980[78,79].

There are, however, some cases where the initial perception of a disease and assigning of a certain name and classification was logical and scientific, but in light of subsequent experience was revealed to be unproductive. The authors believe that in such cases we have to consider whether the disease's name and classification may not be an influencing factor in our failure to treat it effectively and constitute the same cognitive trap that Dr. Tilson was concerned with in case of AAA[59]. The authors suggest that the terms such as "chronic rejection", "transplant obliterative arteriopathy", "transplant arteriosclerosis", or any other of the many technical names listed above, which contain a connotation, directly or indirectly, of alloimmune and rejection nature of the late graft arterial cell proliferation, are exactly what constraints our ability to search for alternative causes of the complication. The impact of these terms is reinforced by the circumstances that arterial neointimal pathology occurs in transplanted organs and is always preceded by acute rejection.

In the 2007 analysis[80] we pointed out that cells in arterial *tunica intima*, and particularly in human coronary arteries, always have a genetic regulation for proliferation and formation of the multilayer cell compartment, which was termed "normal diffuse intimal thickening" or "benign intimal hyperplasia"[81-83]. Furthermore, this multilayer cellular compartment is invariably formed in *tunica intima* of coronary artery in all human hearts, reaching 25-30 cell layers in all young adults. Subsequently, this proliferative morphogenesis slows down to the extent that it maintains the same dimension/morphology for life in most people.

We did not discover this morphogenesis: it was described by Richard Thoma of Heidelberg, Germany, the founder of modern vascular pathology, in 1883[84] and in his subsequent publications[85-88]. The ontogenesis of human coronary artery with an emphasis on the *tunica intima* was described in 1923 by the Russian scientist Kapitoline Wolkoff[89]. In modern times, the ontogenesis and morphology of human coronary arteries were described in great detail in 2002 in the work of Japanese scientists led by Yutaka Nakashima[90]. Notably, the above publications appeared in Virchows Archive, the leading journal in human pathology.

As we can see, this normal morphogenesis of human epicardial coronaries was discovered and confirmed by world-recognized experts, including a detailed description published in the leading journal on human pathology, the Virchows Archive. Nevertheless, numerous studies continue to operate on the wrong assumption that the normal human coronary arterial intima is always an "ideal" single-layer endothelium compartment, and advocate this incorrect perception in the reputed medical journals, e.g.,[91-96], including the Transplantation Journal, e.g.,[53].

Furthermore, as more observations of transplant cardiac arteriosclerosis are reported, it has become evident that the arterial pathology in transplanted hearts is morphologically indistinguishable from coronary pathology due to different causes unrelated to transplantation, e.g., from the pathology of nonatherosclerotic coronary artery disease causing sudden death (Figure 2).

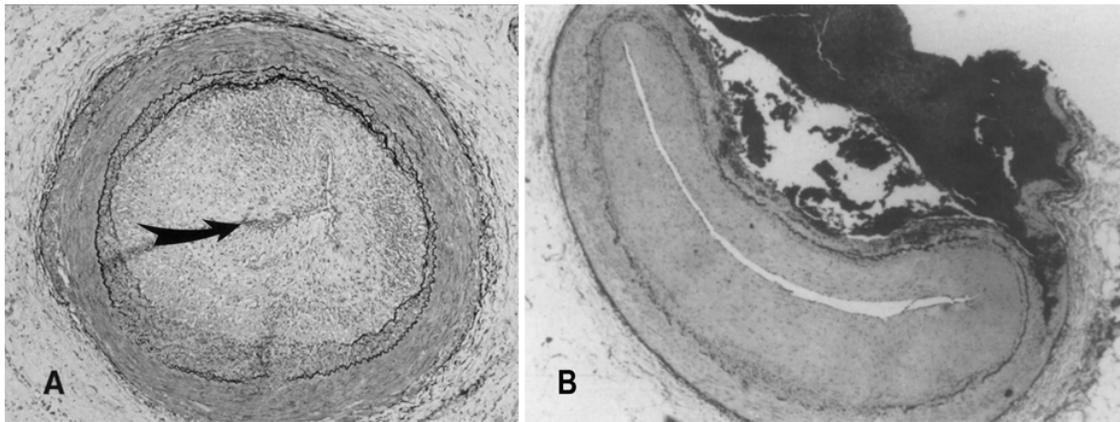

Figure 2. A – distal left anterior descending coronary artery from a cardiac allograft that had been in place for 4 years in a 5-year-old female who underwent retransplantation. Diffuse, fibrocellular intimal proliferation caused virtual total occlusion, with only a slit-like lumen remaining (arrow). Movat pentachrome stain, from[97] with permission of the publisher. B – right coronary artery occlusion in a 55 year-old woman who died from sudden death. Movat pentachrome stain, from[98] with permission of the publisher.

Surprisingly, these similarities were not perceived as a paradox, and neither was the fact that the above cases of pathologic intimal thickening are also indistinguishable from normal arterial morphology of certain segments of human[90] and animal[99] coronary arteries (for a review see[31]).

In a recent analysis we emphasize again that different cases of human coronary pathology due to a variety of identified causes (e.g., alloimmunity, mechanical stress, smoking, high blood pressure, misbalanced level of blood lipoproteins, repeated courses of radiation, lack of physical activity, etc.) or without any known causes (i.e., sudden coronary death due to idiopathic coronary arteriosclerosis), appeared to be microscopically identical or very similar to each other[31]. In the same publication we raised questions and suggested possible answers, both of which are important for the present analysis and can be reiterated verbatim:

"Let us appeal to common sense and ask a question: What kind of primary stimulatory signals could be shared between strong pathophysiologic reactions such as alloimmunity and mechanical stress, on one hand, and the subtle beginning of coronary atherosclerosis? Or between alloimmunity, mechanical stress, and the absence of any tangible signals at all? The answer is none; yet in all cases an identical specific morphogenesis is directed (Figures 1a, 6 and 5a in[31]). For the sake of clarity, let us visualize the above questions:

alloimmunity→GVD→neointimal formation;

stent→mechanical stress restenosis→neointimal formation;

smoking→peripheral artery diseases→neointimal formation;

(high blood pressure, misbalance level of blood lipoproteins, lack of physical activity, etc.)→coronary atherosclerosis→neointimal formation;

absence of any tangible etiology and cause→idiopathic intimal hyperplasia→neointimal formation;

etc.

The suggestion that identical pathologic pattern formation can be directed by signals that are extremely different in nature and magnitude is very improbable."[31]

By implying that alloimmunity or mechanical stress are the main stimuli directing arterial neointimal morphogenesis we go against scientific logic, because we routinely observe identical deadly neointimal formation without any detectable stimuli at all[98,100,101]. How can we find a way out of this enigma?

**Tissue competence, self-organization and endogenous control of gene expression**

Let us try to find what is common for all these different purported causes of pathologic neointimal formation. The answer may seem surprising: all the suggested causes are of exogenous nature with respect to the coronary artery, i.e. they act from the outside of the coronary wall. However, this is hardly a revelation.

In the monograph "Pathogenesis of coronary artery disease" (1969), Meyer Friedman writes:

"… an artery responds to almost any sort of physical or chemical injury to which it may have been exposed with a hyperplasia or replication of its surviving or still-intact remnants. Thus, whether the injury is induced by transplantation, by needle puncture, by freezing, by heat, by exposure to electron radiation, by induced hypertension, by cholesterol infiltration, or by pressor amines, the typical response on the tunic affected is a hyperplastic one"[102].

If we posit that exogenous signals of any nature lead an identical proliferative response, i.e. that different exogenous insults direct the same hyperplasic morphogenesis in the arterial intima, this may seem to go against the principle of theoretical parsimony.

However, the above notion can be coherent in light of the phenomenon called tissue competence[103]. A tissue that is classified as *competent* or *poised* can respond only in a specific way to a variety of non-specific stimuli[104]. This response is thought to be not an accident but rather a consequence of expression of *preexisting* genetic regulation. Stuart Kauffman writes in "The origins of order: Self-organization and selection in evolution", 1993:

"A particularly striking fact that emerged from the initial attempts to isolate the normal inducer molecules for neurectoderm development is that cell types are poised among only a few alternative pathways of differentiation. A variety of compounds, including pure chemical substances, even changes in pH and pure water, were found to induce embryonic ectoderm to differentiate into neurectoderm. Thus the critical conclusion is that the ectoderm is, as noted, poised between only two alternative pathways and can be triggered to differentiate into neurectoderm by a variety of normal and abnormal stimuli."[105].

The above arguments are cited here to support the following suggestion: resident cells in coronary *tunica intima* always possess the genetic regulation that can govern transformation of normal tunica intima into pathological neointimal formation. This statement is also applicable to the *tunica intima* of any artery, as well as to the tunica intima of veins, since neointimal failure of the venous conduit in bypass surgery is notorious.

The above means that cells in coronary *tunica intima* **always** possess a genetic regulation, which can direct transformation of a normal design of coronary *tunica intima* into pathology[3], e.g. into the chronic vascular rejection pathology. However, in naïve hearts and in some of the transplanted hearts, these genes are not expressed, in other words the genetic regulation is "turned off" (suppressed), preventing excessive pathologic neointimal formation or chronic rejection. It is the basic mechanism that functions in all cells, from bacterial to mammalian: the endogenous control of gene expression[106, 107].

---

[3] Previously, we referred to the evolutionary developmental mechanisms to explain the origins of this morphogenesis, and proposed a model in which the properties of blood-tissue interface act as the endogenous control of genetic regulations[80]. The above conjecture is omitted in this writing to avoid distracting readers from the main argument of the present analysis.

Let us return to the subject, i.e., chronic rejection of the transplanted heart. Based on the presented evidence, we can reasonably posit that cells in coronary arteries of all human hearts possess genetic regulation that can transform normal *tunica intima* of coronaries into the same deadly pathology, which in this case manifests itself as obliterative arteriopathy after heart transplantation. The same is true for other coronary pathologies: in-stent restenosis, coronary occlusion in Kawasaki disease, idiopathic intimal hyperplasia causing sudden coronary death (for examples of diseases see [31]). By the same token, we can reasonably assert that this genetic regulation is controlled (or suppressed) in some transplanted or stented hearts, effectively preventing allograft vasculopathy (chronic vascular rejection) or in-stent restenosis.

Unfortunately, we do not yet know what genetic regulation maintains normal coronary morphology in some transplanted and stented hearts, as well as in non-diseased hearts. But we do know for sure that this regulation can be altered in chronic rejection after organ transplantation, as well as in other diseases, by numerous stimuli, suggesting that they logically are non-specific triggering factors and not the cause of pathology. The real cause is the loss of control of the genetic regulation that governs normal coronary morphogenesis. This controlling regulation can also fail without any apparent external case, as in idiopathic intimal hyperplasia causing sudden coronary death[98,101,108].

To put it simply: there is an as yet poorly understood genetic regulation that maintains normal coronary morphology, whose mechanism must be investigated to address this dire public health crisis[54]. We also know from numerous observations that this genetic regulation can be altered by variety of non-specific factors. However, instead of studying this common target that all these non-specific factors act upon, we devote all our efforts to studying numerous non-specific factors in isolation.

Of course, alloantigens and persistent activation of host immune responses against donor tissues are evident in chronic rejection. Nevertheless, to focus primarily on immune triggering signals, constitutes, in our opinion, an unproductive approach, or what Lahmann and Joner recently described, in regard to in-stent restenosis, as "chasing out tails in search of a solution"[109].

**Summary and conclusions**

So long as we continue using the terms "chronic rejection" and "alloimmunity" we are encouraging the study of the allograft vasculopathy chiefly from the perspective of alloantigens and activation of the recipient immune system.

By considering all cases of neointimal pathologies as phenomena of the same nature, we suggest that the allograft vasculopathy is not an alloimmune phenomenon and not a form of rejection. This gives us a different perspective on the failure of otherwise highly effective immunophilins to treat chronic rejection. The failure to prevent/treat chronic vascular rejection lies not with the immunophilins but rather with our alloimmune theory of the pathology. We also suggests that in the case of in-stent restenosis, it is not mechanical stress that induces neointimal formation in restenosis; otherwise all patients who underwent balloon dilatation or bare metal stenting should have developed restenosis, which we know is not the case.

Why do some patients today still develop restenosis after stenting with the best stent designs, while others do not? This question mirrors the one asked by Thomas E. Starzl,

known as the "father of modern transplantation":

"Why did some people reject their transplants while others didn't, and others didn't seem to need the drugs at all?"[110].

There is as yet no answer to this question. The authors hope that the above analysis helps to facilitate a fresh look at the problem.

First, we have to reconsider how we name and classify a disease. The name "chronic vascular rejection" and other alternative names, i.e. transplant atherosclerosis and more than a dozen others (see above), even the "neutral" GVD, all carry the connotation to alloimmunity and rejection. But we can now plausibly infer from fifty years of exhaustive studies and numerous lost lives that all approaches to allograft vasculopathy within the alloimmunity rejection model are likely fruitless and doomed to failure. We must step out of the "alloimmunity rejection box".

We have to think carefully whether the name we give to a disease could affect its study and our ability to find a cure, and if we gave the wrong name to a disease, we have to change it, because, to quote from the same passage by Nietzsche again: "…How foolish it would be to suppose that one only needs to point out this origin and this misty shroud of delusion in order to destroy the world that counts for real, so-called "reality". We can destroy only as creators."[60] Today, calls for such linguistic reexamination in medical science, and particularly concerning vascular diseases, are not unusual, e.g.,[111], but they have not been widely acted upon.

Taking into account 1) all known facts regarding the mysterious arterial intimal hyperplasia in transplanted organs and 2) our tragic failure to prevent and treat allograft vasculopathy using all available knowledge on alloimmunity and rejection, we must finally disassociate the former from the latter. A good way to start this uncomfortable process is to change the words we are using; particularly, the words we chose to name the disease. By correcting our linguistic notation, i.e., names and systematic classification of the disease, we can avoid potential pitfalls associated with the use of incorrect language in science in general[112], and particularly in medical science[59] and in the case of late allograft deterioration[113]. It is worth noting that the customary appeal to the famous line of Shakespeare "What's in a name? That which we call a rose by any other name would smell as sweet" is viewed very critically by many medical practitioners and scientists when it comes to the names of diseases[114-117].

We conclude as we began: "Let's pause between our words … so that the meaning of the word just spoken, sounds a clearer echo in our heads."[1].

**Dedication**

The authors dedicate this work to the memory of Alla Subbotin (08/10/1945 – 08/03/2023), beloved wife, mother and friend (Figure 3). Alla witnessed the moment of conception of this analysis, then encouraged us to write it together and followed the writing almost to the last lines, when we did not know how to finish it. Now we know.

**We love and admire you. Our grief has no bounds.**

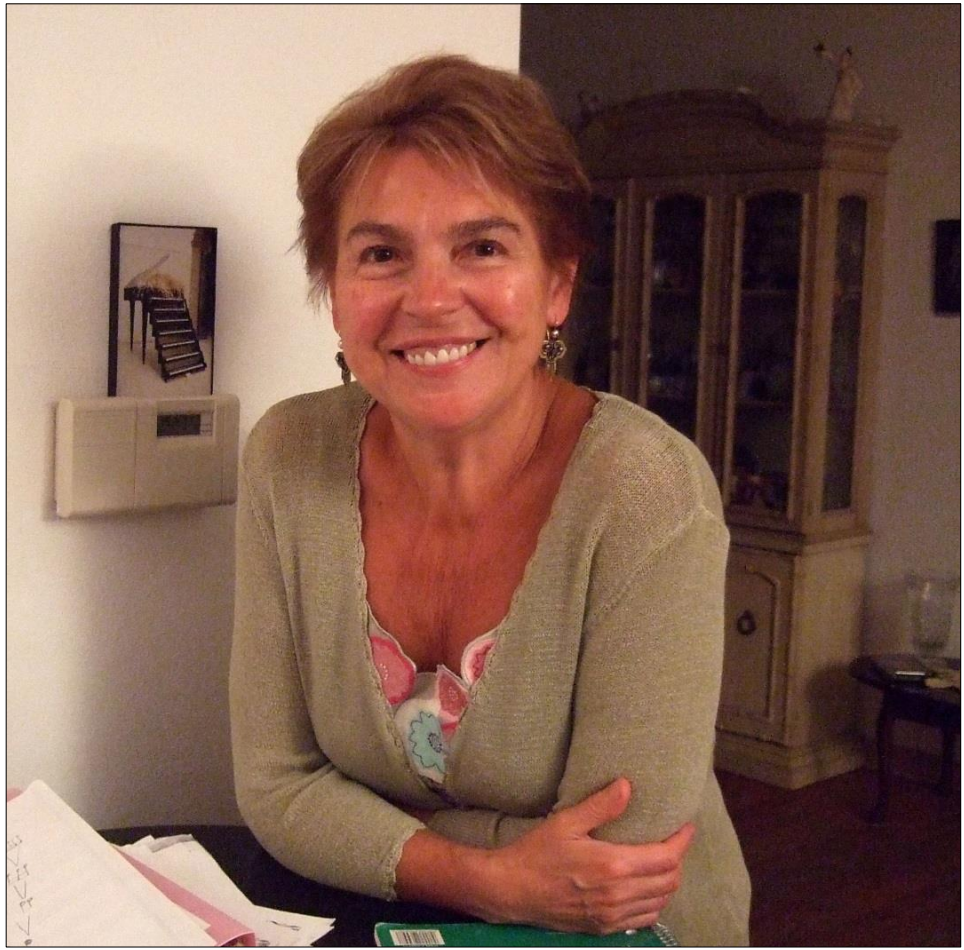

Figure 3. Alla Subbotin.